\documentstyle[12pt]{article}

\def\hybrid{\topmargin 10pt  \oddsidemargin 0pt
      \headheight 0pt   \headsep 0pt
      \textwidth 6.25in 
      \textheight 8in 
      \marginparwidth .875in
      \parskip 5pt plus 1pt   \jot = 1.5ex}

\hybrid
\begin{document}
\def\eeqa{\end{eqnarray}}
\sloppy
\renewcommand{\arraystretch}{1.6}
\newcommand{\be}{\begin{equation}}
\newcommand{\eq}{\end{equation}}
\begin{titlepage}
\begin{center}
\hfill HUB-EP-97/26\\
\hfill hep-th/9704147\\
\vskip .6in
{\large\bf Black Holes In $N=2$ Supergravity Theories And Harmonic Functions}
\vskip .5in
{W. A. Sabra}\footnote{e:mail: sabra@qft2.physik.hu-berlin.de}\hfill
\vskip 0.5cm
\hfill

{\em Humboldt-Universit\"at zu Berlin,\\
Institut f\"ur Physik,\\ 
Invalidenstra\ss e 110,\\
D-10115 Berlin, Germany}\\
\end{center}
\vskip .5in
\begin{center} {\bf ABSTRACT } 
\end{center}
\begin{quotation}\noindent
We present dyonic BPS static black hole solutions 
for general $d=4$, $N=2$ 
supergravity theories coupled to vector and hypermultiplets. 
These solutions are 
generalisations of the spherically symmetric Majumdar-Papapetrou 
black hole solutions of
Einstein-Maxwell gravity and are completely characterised by  
a set of constrained harmonic functions.
In terms of the underlying special geometry, these harmonic functions are 
identified with the imaginary part
of the holomorphic sections defining the special K\"ahler manifold and the
metric is expressed in terms of the symplectic invariant K\"ahler potential. 
The relations of the holomorphic sections to the harmonic functions 
constitute the \lq\lq{\it generalised stabilisation equations}" for 
the moduli fields.
In addition to asymptotic flatness, the harmonic functions are also 
constrained by the requirement that the  K\"ahler connection of the underlying 
Hodge-K\"ahler manifold has to vanish in order to obtain static solutions.
The behaviour of these solutions near the horizon is also explained.
\end{quotation}
\end{titlepage}
\vfill
\eject
\newpage
\section{Introduction}
There has been lots of interest as well as progress in the study 
of BPS black holes in ungauged four-dimensional $N=2$ supergravity 
theories 
coupled to vector and hypermultiplets \cite{fks}-\cite{fkg}. 
The underlying special 
geometry structure of the $N=2$ theory has been a very useful tool in 
the analysis of these black holes\footnote{For a review on $N=2$ 
supergravity and 
special geometry see for example \cite{fre}}. 
In a general $N=2$ supersymmetric 
theory, the mass of a BPS state 
which breaks half of the supersymmetry is given in terms of the modulus 
of the central charge of the underlying $N=2$ 
supersymmetry algebra \cite{CDFP}. The central charge  
is a function of the scalars of 
the theory as well as the electric 
and magnetic charges which correspond to the $U(1)^{n+1}$ gauge symmetry in 
a theory with $n$ 
abelian vector multiplets.\footnote{The extra $U(1)$ is due to the 
graviphoton.} 
Moreover, the black hole ADM mass in the $N=2$ theory is governed by the
value of the central charge at infinity and thus depends on 
the electric and magnetic charges as well as the asymptotic values of the 
scalar fields. 

The recent work of Ferrara, Kallosh and Strominger 
\cite{fks,s,fk} provided an algorithm for the macroscopic determination 
of 
Bekenstein-Hawking entropy \cite{bh} for the $N=2$ extremal black 
holes. One simply extremizes the central charge at fixed values of   
electric and magnetic charges and the extremum value $Z_h$ of the central 
charge 
gives the entropy $S,$ which is quarter of the area $A$ of the 
horizon, 
via the relation \cite{fk}    
\be 
S={A\over 4}=\pi\vert Z_{h}\vert^2  
\eq   
     
Moreover, the horizon acts as an attractor for the scalar fields. 
This means that the values of the moduli at the  
horizon are fixed and completely 
independent of their values at spatial infinity. 
In general the fixed values for the scalars at the horizon are 
those which extremize the central charge. The near horizon physics 
also depends on topological data in the cases where 
the $N=2$ supergravity models are 
obtained from compactifying type-II string theories on Calabi-Yau three-folds. 
The prepotential of these theories depend on the classical intersection 
numbers, Euler number and rational instanton numbers. 
Therefore, for these models, the  central charge and  
the entropy depend on the electric and magnetic charges and 
the topological data of the Calabi-Yau manifold \cite{qe}. 

The above results were obtained by making use of the supersymmetry 
transformation rules for the gravitino and 
the gauginos in the ungauged bosonic part of 
$N=2$ supergravity coupled to vector multiplets, where it was also 
assumed that the hypermultiplets take constant values. 
The condition for unbroken supersymmetry near the horizon is the statement 
that the holomorphic covariant derivative of the central charge 
(related to the 
graviphoton field strength) must vanish. At this point, 
one obtains equations which relate the moduli to the magnetic
and electric charges and other discrete parameters.
Using the extremization procedure 
of the central charge \cite{fk}, the entropy formulae, derived from the
explicit solutions, for $N=4$ and $N=8$
extreme black hole \cite{mc} were obtained. The entropy of $N=2$ black 
holes has been studied 
for particular models both at the classical, perturbative and
non-perturbative level \cite{ksw}-\cite{me}. It should be mentioned that 
while the entropies for 
$N=4, 8$ black
holes are unique and are determined in terms of quantised magnetic and
electric charges, those for the $N=2$ theories depend on the specific details
of the underlying special K\"ahler manifold.

A special class of black hole solutions, the so called double extreme 
black holes, is obtained if one assumes that the moduli fields 
take the same fixed value from the horizon to spatial infinity \cite{ksw}. 
For these solutions, the central charge is constant 
everywhere and thus the black hole ADM 
mass coincides with the Bertotti-Robinson mass just as in the case of 
pure Einstein-Maxwell theory. The metric of the double extreme black 
hole is of the extreme Reissner-Nordstr{\o}m form and can be expressed as
\begin{eqnarray}
ds^2 &=&\Big(1+{{\sqrt{A/ 4\pi}}\over r}\Big)^{-2}dt^2-
\Big(1+{{\sqrt{A/ 4\pi}}\over r}\Big)^{2}(d\vec x)^2\nonumber\\
&=&\Big(1+{\vert Z_{h}\vert\over r}\Big)^{-2}dt^2-
\Big(1+{\vert Z_{h}\vert\over r}\Big)^{2}(d\vec x)^2,\label{rn}
\end{eqnarray}
where $r\equiv\sqrt {\vec{x}\cdot\vec{x}},$
and $\vec x$ is the position vector in flat three-dimensional space 
referred to as the 
background space.
The mass of the black hole is defined by 
\be
g_{tt}=(1-{2M_{ADM}\over r}+\cdots )
\eq
and from (\ref{rn}) one obtains 
\be
M_{ADM}={\sqrt{A\over 4\pi}}=\vert Z_{h}\vert.
\eq
Therefore, the double extreme black hole solution is, in principle, 
known for all $N=2$ supergravity theories and is completely specified in 
terms of the central charge of the underlying $N=2$ supersymmetry algebra.  
Thus if we denote the
covariantly holomorphic sections by 
$(L^I, M_I)$\footnote{These quantities are defined in the next section}, then
the central charge $Z$ can be expressed by
\be
Z=M_Ip^I-L^I q_I,
\eq 
where $q_I$ and $p^I$ are the electric and magnetic charges. The double
extreme black hole is then completely determined by 
\be
Z_h= M_{hI}p^I-L_h^I q_I,
\eq
where $(L_h^I, M_{hL})$ are the values of the covariantly holomorphic 
sections evaluated at the near horizon, and are given in terms of 
the \lq\lq {\it stabilisation equations}"
\begin{eqnarray}
i(Z_h\bar L^I-\bar Z_h L^I)&=&p^I,\nonumber\\
i(Z_h\bar M_I-\bar Z_hM_I)&=&q_I,
\label{deb}
\end{eqnarray}
which are obtained from the
extremization of the central charge.

The simplicity of the double extreme black holes comes from 
restricting the moduli fields to constant values everywhere. 
However, near the horizon, both of the extreme and 
the double extreme dyonic BPS black holes lose all their scalar hair 
and depend only on 
conserved charges and discrete parameters 
corresponding to gauge symmetries and topological data.
The near horizon can be approximated by 
the Bertotti-Robinson type metric where the area of the black
is interpreted as the mass of the Bertotti-Robinson universe \cite{br} 
\begin{eqnarray}
ds^2& =& {r^2\over M^2_{BR}}dt^2-{M^2_{BR}\over r^2}(d\vec 
x)^2\nonumber\\
& =&{4\pi r^2\over A}dt^2-{A\over 4\pi r^2}(d\vec x)^2\nonumber\\
&=&{r^2\over\vert Z_{h}\vert^2}dt^2-{\vert Z_{h}\vert^2\over r^2}
(d\vec x)^2.
\end{eqnarray}

The Bertotti-Robinson metric plays a special role 
in Einstein-Maxwell gravity in the sense that it can be considered 
as an alternative, maximally supersymmetric vacuum state. 
The extreme Reissner-Nordstr{\o}m 
metric is a soliton which breaks half of the supersymmetry and interpolates 
between two maximally supersymmetric configurations; 
the trivial flat metric vacuum and the Bertotti-Robinson 
vacuum \cite{G}. Maximally supersymmetric configurations are of course 
those where the full $N=2$ supersymmetry is restored. 
For BPS black holes in  $N=2$ supergravity theories with vector multiplets, 
the near horizon metric, 
as for the pure supergravity case, is still of the Bertotti-
Robinson type. The additional feature is that the 
unbroken supersymmetry near the horizon restricts the moduli to 
fixed discrete values independent of their initial values at spatial infinity.

General extreme purely magnetic $N=2$ black hole solutions were 
derived in \cite{fks}. Also extreme solutions were constructed for the 
so called axion-free $STU$ model associated with the special K\"ahler 
manifold ${SU(1,1)\over U(1)}\times 
{SO(2,2)\over SO(2)\times SO(2)}$in 
\cite{ber} and for supergravity models based on the 
minimal coupling manifolds 
${SU(n, 1)\over SU(n)\times U(1)}$ in \cite{me}. 
In the construction of \cite{me} it was realised 
that the general extreme black hole solutions 
can be constructed from a set of constrained 
complex harmonic functions. Later in \cite{mea}, 
general extreme static black holes solutions were 
derived for an arbitrary $N=2$ supergravity theory 
coupled to vector and hypermultiplets.  

In this work, our purpose is to study 
general dyonic BPS extreme black hole solutions, 
for non-constant complex values of the 
moduli, and for an arbitrary $N=2$ supergravity model coupled to 
vector and hypermultiplets. In deriving these black hole solutions, 
we obtain \lq\lq{\it  generalised stabilisation equations}" 
expressing the values of the moduli 
in terms of the charges at any point in space-time.
Using these explicit solutions, the near horizon physics 
namely, the results of \cite{fk} are rederived. 
As a simple illustration of our results, 
we derive the known dyonic Reissner-Nordstr{\o}m black hole solutions 
of Einstein-Maxwell gravity using the special geometry 
formulation of pure $N=2$ supergravity theory.
Part of the results of this work were briefly presented in \cite{mea}.

$N=2$ supergravity theories with vector and hypermultiplets are 
fully defined in terms of the special geometry of the manifold 
spanned by the scalars of the
vector multiplets. In the analysis of the black hole solutions, we will use 
the symplectic formulation of special geometry which does not 
depend on the existence of a holomorphic prepotential. 
This work is organised as follows. In the next section, some 
basics of special K\"ahler geometry and $N=2$ supergravity coupled to vector 
multiplets are reviewed where we display formulae 
relevant for our later discussions.
The extreme black hole solutions with vector multiplets are then considered
in section three. Section four contains a re-derivation of the 
Reissner-Nordstr{\o}m solution of pure 
Einstein-Maxwell theory using the framework of special geometry.
In section five, we use our general solutions 
and verify that the near horizon is approximated by a Bertotti-Robinson
universe, where the scalars are always given in terms of the charges, 
and with values which extremise the
central charge. The central charge evaluated at the horizon gives the
Bekestein-Hawking entropy. The last section contains  a summary and 
a discussion on how our results can be extended to the 
construction of stationary solutions.  
\section{Special Geometry and $N=2$ Supergravity}
In recent years, special geometry has emerged as an essential structure 
in the study of $N=2$ supergravity theories, the vacuum structure of 
superstrings, topological field theories and mirror symmetry in 
Calabi-Yau three-folds. 
More recently, special geometry 
provided a useful tool in the study of the quantum moduli space and 
obtaining exact solutions of low energy effective actions 
for rigid and local $N=2$ supersymmetric Yang-Mills theories \cite{fre}.

The concept of special K\"ahler geometry was first introduced to the 
physics literature in the analysis of $N=2$ supergravity models coupled 
to vector multiplets \cite{sspecial}. There special K\"ahler manifolds were 
defined by the
coupling of $n$  vector multiplets to 
$N=2$
supergravity. The complex scalars $z^i$ of the 
$N=2$ vector multiplets coupled to supergravity 
are coordinates of a special K\"ahler manifold. 
An intrinsic definition of special K\"ahler 
geometry in 
terms of symplectic bundles was later 
given \cite{cc} in connection with the geometry of the moduli of 
Calabi-Yau 
spaces where special K\"ahler manifolds were associated with the moduli 
space 
of the K\"ahler or complex structure. Also a coordinate-independent 
description was given in \cite{ccc,cccc}, where special geometry was 
derived
from the constraints of the extended $N=2$ supersymmetry in the 
non-linear sigma models associated with an arbitrary number $n$ of vector 
multiplets 
of a four dimensional supergravity. 
Special K\"ahler manifolds 
are K\"ahler-Hodge manifolds, with an additional constraint on the curvature 
\cite{sspecial}
\be
R_{ij^\star k l^\star}=g_{ij^\star}g_{k l^\star}+g_{il^\star}g_{k 
j^\star}-
C_{ikp}C_{j^\star l^\star p^\star}g^{p p^\star},
\label{monster}
\eq
where $g_{ij^\star}=\partial_i\partial_{j^\star}K,$ is the K\"ahler metric 
with $K$ the K\"ahler potential and $C_{ijk}$ is a completely symmetric 
covariantly holomorphic tensor. K\"ahler-Hodge manifolds are 
characterised by a $U(1)$ bundle 
whose first Chern class is equal to the K\"ahler class. 
This implies that, locally, the $U(1)$ connection can be represented by
\be
Q=-{i\over2}(\partial_iK dz^i-\partial_{i^\star}Kd\bar z^{i^\star}) 
.\eq
An intrinsic definition of special K\"ahler manifold can be given 
\cite{c}-\cite{cccc} in terms 
of a flat $2n+2$ dimensional symplectic bundle over the K\"ahler-Hodge 
manifold, with the covariantly holomorphic sections
\begin{eqnarray}
V&=&\pmatrix{L^I\cr M_I}, \qquad I=0,\cdots,n
\nonumber\\
D_{i^\star}V&=&(\partial_{i^\star}-{1\over2}\partial_{i^\star}K)V=0,
\label{saigon}
\end{eqnarray}
obeying the constraints \cite{vp}
\be
i<V\vert\bar V>=i(\bar L^I M_I-L^I\bar M_I)=1,\qquad <V, U_i>=0
\label{berlin}
\eq
where the symplectic inner product is understood to be taken with 
respect to 
the metric 
$\pmatrix{0&-1\cr 1&0},$ and 
\be
U_i=D_i V=(\partial_{i}+{1\over2}\partial_{i}K)V=
\pmatrix{f_i^I\cr h_{iI}}.
\eq
In general, $D_i$ is the covariant derivative with respect 
to the 
Levi-Civita 
connection and the connection $\partial_iK$. Thus, for a generic field 
$\phi^i$ which transforms under the K\"ahler 
transformation, $K\rightarrow K+f+\bar f$, by the $U(1)$ transformation
$\phi^i\rightarrow e^{-({p\over2}f+{{\bar p}\over2}{\bar f})}\phi^i,$ 
we have
\be
D_i\phi^j=\partial_i\phi^j+\Gamma^j_{{ik}}\phi^k+{p\over2}{\partial_i 
K}\phi^j.
\eq
One also defines the covariant derivative $D_{i^\star}$ in the same way but 
with $p$ replaced with $\bar p.$
The sections $(L^I, M_I)$ has the weights $p=-\bar p=1$ and $C_{ijk}$ has the
weights $p=-\bar p=2$.

In general, one can write
\begin{eqnarray}
M_I&=&{\cal N}_{IJ}L^J,\nonumber\\
h_{iI}&=&{\bar{\cal N}}_{IJ}f_{i}^J.
\label{warum}
\end{eqnarray}
The complex symmetric $(n+1)\times (n+1)$ matrix $\cal N$ 
encodes the couplings of the vector 
fields in the corresponding $N=2$ supergravity theory. 

It can be shown \cite{special}-\cite{sspecial} that the condition 
(\ref{monster}) can be obtained from the integrability conditions on the 
following differential constraints
\begin{eqnarray}
D_iV &=& U_i,\nonumber\\
D_iU_j & = & iC_{ijk}g^{kl^\star}{\bar U}_{l^\star},\nonumber\\
D_i{\bar U}_{j^\star} & = &g_{ij^\star}{\bar V},\nonumber\\
D_i{\bar V} & =& 0.
\label{set}
\end{eqnarray}
It is well known that the constraints (\ref{set}) 
can in general be solved in terms of a holomorphic function of degree 
two \cite{sspecial}. However, there exists symplectic sections for 
which such a holomorphic function does not exist. This, for example, 
appears in the study of the 
effective theory of the 
$N=2$ heterotic strings \cite{CDFP}. Thus it is more natural to use the 
differential constraints 
(\ref{set}) as the fundamental equations of special geometry.

The K\"ahler potential can be constructed in a symplectic invariant 
manner as
follows. Define the sections $\Omega$ by
\be
V=\pmatrix{L^I\cr M_I}=e^{K\over2}\Omega=e^{K\over2}\pmatrix{X^I\cr 
F_I}
.\eq
It immediately follows from (\ref{saigon}) that $\Omega$ is 
holomorphic; 
\be
\partial_{i^\star}X^I=\partial_{i^\star}F_I=0.
\eq
Using ({\ref{berlin}), one obtains 
\begin{eqnarray}
K&=&-\log\Big(i<\Omega\vert\bar\Omega>\Big)\nonumber\\
&=& -\log\Big[i(\bar X^IF_I-X^I\bar F_I)\Big].
\end{eqnarray}

Exploiting the relations (\ref{berlin}),(\ref{set}) and (\ref{warum}), 
the 
following 
symplectic expressions can be derived for the K\"ahler metric and 
$C_{ijk}$
\begin{eqnarray}
g_{ij^{\star}}&=&
-i<U_i\vert\bar U_{j^\star}>=
-2f_i^I \mbox{Im}{\cal N}_{IJ}\bar f_{j^\star}^J\\
C_{ijk}&=&<D_iU_j, U_k>.
\end{eqnarray}
For our purposes, it is also useful to display the following relations 
\be
g^{ij^\star}f_i^I{\bar f}_{j^\star}^J=-
{1\over2}(\mbox{Im} {\cal N})^{-1IJ}-{\bar L}^IL^J,
\qquad \mbox{Im}{\cal N}_{IJ} L^I{\bar L}^J
=-{1\over2}.
\label{dirac}
\eq

It should be mentioned that the dependence of the gauge couplings
on the scalars characterising homogenous special K\"ahler manifolds 
of $N=2$ supergravity 
theory can also be determined from the knowledge of the corresponding
embedding of the isometry group
of the scalar manifold into the symplectic group 
\`a la Gaillard and Zumino \cite{gz, sym}.

The $N=2$ supergravity action includes 
one gravitational, $n$ vector and hypermultiplets. However, 
for our purposes, we will neglect 
the hypermultiplets in what follows and assume that these fields are 
constants. In this case, the bosonic $N=2$ action is
given by
\be
S_{N=2}= \int \sqrt{-g} d^4x \Big(-{1\over2}R+g_{i {{j}^\star}}
\partial^{\mu} z^i \partial_{\mu} \bar z^{{j}^\star}  
+ i\left(\bar {\cal N}_{I J} 
{F}^{- I}_{\mu \nu}
{F}^{- J {\mu \nu}}
\, - \,
{\cal N}_{IJ} {F}^{+ I}_{\mu \nu}
{ F}^{+ J {\mu \nu}}\right)  
\label{bose}
\eq
where 
\be
{F}_{\mu\nu}^{I\pm}={1\over2}\Big({F}_{\mu\nu}^{I}\pm 
{i\over2}\varepsilon_{\mu\nu\rho\sigma}{F}^{\rho\sigma I}\Big)
\eq

The important field 
strength combinations which 
enter the chiral gravitino and gauginos supersymmetry transformation 
rules are given by
\begin{eqnarray}
T^{-}_{\mu\nu}&=& M_I F^I_{\mu\nu}-L^I G_
{I \mu\nu}=2i(\mbox{Im} {\cal N}_{IJ})L^I F^{J-}_{\mu\nu}\\
G_{\mu\nu}^{-i}&=&-g^{ij^{\star}}{\bar f_{j^\star}^I} (\mbox{Im}{\cal 
N})_{IJ}F^{J-}_{\mu\nu}.
\end{eqnarray}
The supersymmetry transformation for the chiral gravitino 
$\psi_{\alpha\mu}$
and gauginos $\lambda^{i\alpha}$ in a bosonic background of 
$N=2$ supergravity
are given by
\begin{eqnarray}
\delta\,\psi_{\alpha\mu} &=& \nabla_\mu \epsilon_\alpha -
\frac{1}{4} T^-_{\rho\sigma} \gamma^{\rho\sigma}
\, \gamma _\mu \, \varepsilon_{\alpha\beta}\epsilon^\beta \, 
\label{grtrans},\\
\delta\lambda^{i\alpha} & = & 
i\partial_\mu z^i\gamma^\mu\epsilon^\alpha 
+ G^{-i}_{\rho\sigma}\gamma^{\rho\sigma}
\epsilon^\beta\varepsilon ^{\alpha\beta}
\label{gatrans}
\end{eqnarray}
where $\epsilon_\beta$ is the chiral supersymmetry parameter, 
$\varepsilon^{\alpha\beta}$ is the $SO(2)$ Ricci tensor and 
the space-time covariant derivative $\nabla_{\mu}$ also 
contains the K\"ahler connection
\be
Q_\mu= -{i\over2}\Big(\partial_i K\partial_\mu z^i-
\partial_{i^\star} K\partial_\mu {\bar 
z}^{i^\star}\Big),
\eq
Therefore we have 
\be
\nabla_\mu \epsilon_\alpha=(\partial_\mu-{1\over4}w^{ab}_\mu\gamma_{ab}+
{i\over 
2}Q_\mu)\epsilon_\alpha 
\eq
where $w^{ab}_\mu$ is the spin connection. 

The mass of a BPS state breaking half of the supersymmetry is given in 
terms 
of the central 
charge $Z$ of the theory is defined by \cite{CDFP}
\be
M_{BPS}=\vert Z\vert^2=\vert M_I p^I-L^I 
q_I\vert^2=e^{K}
\vert F_I p^I-X^I q_I\vert^2
\eq
where the electric and magnetic charges are defined by
\begin{eqnarray}
q_I &=&\int_{S^2_\infty}  G_{I \mu\nu}dx^\mu\wedge 
dx^\nu\nonumber\\
p^I &=&\int_{S^2_\infty} F^{I}_{\mu\nu}dx^\mu\wedge dx^\nu
\end{eqnarray}
where ${S^2_\infty}$ is the two-sphere at infinity.

\section{General Black Hole Solutions}
In this section we derive general static BPS black hole solutions for 
$N=2$ supergravity theory with an arbitrary number of vector multiplets.
It is known \cite{gh,tod} that static solutions 
admitting 
supersymmetries are given by the 
Majumdar-Papapetrou black holes solutions \cite{mp}. Here we consider 
spherically symmetric 
solutions which can be written in the form
\be
ds^2 =e^{2 U(r)} dt^2 -e^{-2 U(r)} (d\vec{x})^2, \qquad
\eq

In order to find the explicit BPS black hole 
solution 
it is more convenient to use the 
supersymmetry transformations rules of the fermi fields since 
these transformation 
rules
depend linearly on the first derivatives of the bosonic fields and thus
their vanishing provide first order differential equations \cite{fks, fr}.

{From} the conditions of vanishing supersymmetry transformation, $i.e.,$
$\delta\psi _{\alpha\mu} =\delta\lambda^{i\alpha}= 0,$ one obtains, 
for a particular choice of the supersymmetry parameter, the following
first order differential equations \cite{fr, mea}
\begin{eqnarray}
\frac{dU}{dr} & = &-2\mbox{i}
\left(\mbox{Im}{\cal  N}\right)_{IJ}
L^I t^J\frac{e^U}{r^2}
\label{eqone}\\
\frac{dz^i}{dr}&= &-2\mbox{i}g^{ij^\star}{\bar f}^I_{j^\star}
\left(\mbox{Im}{\cal N}\right)_{IJ}\frac{t^J}{r^2} e^U
\label{eqtwo}
\end{eqnarray}
where 
\be
t^I={1\over2}
\Big(p^I-i(\mbox{Im}{{\cal N}^{-1}})^{IK}(\mbox{Re}{\cal N})_{KM}
p^M+i(\mbox{Im}{{\cal N}^{-1}})^{IK}q_K\Big).\eq
The solution of the above equations of course depends on the particular 
model one is considering, $i.e.,$ the choice of the special K\"ahler 
manifold. In what follows we will solve the above
differential equations in a model independent way. 
To begin, consider the first differential equation 
(\ref{eqone}).
This can be rewritten as
\be
{dU\over dr}=-Z{e^U\over r^2}=e^{K\over2}(X^Iq_I-F_Ip^I){e^U\over r^2}
.\label{g}
\eq
Our ansatz for the solution is to take 
\be
e^{-2U}=e^{-K}=i(\bar X^IF_I-X^I\bar F_I),
\label{s}
\eq
which is the most natural choice due to the symplectic invariance of
 the K\"ahler 
potential. This choice enables us to rewrite (\ref{g}) 
in the following form 
\be
{d\over dr}{e^{-2U}}={d\over dr}{e^{-K}}=-2{(X^Iq_I-F_Ip^I)\over r^2}
.\label{gg}
\eq
Differentiating $e^{-K}$ with respect to $r$, we get
\be
{d\over dr}e^{-K}=i({d\bar X^I\over dr}F_I+\bar X^I{dF_I\over dr}-
{dX^I\over dr}\bar F_I-{X^I}{d\bar F_I\over dr})
\label{tan}
\eq
and demanding that our solutions satisfy, 
\be
\bar X^I{dF_I\over dr}-{dX^I\over dr}\bar F_I={d\bar X^I\over dr}F_I-
{X^I}{d\bar F_I\over dr},
\label{ha}
\eq
then from (\ref{gg}) we obtain 
\be
{d\over dr}e^{-2U}=2i\Big({d\bar X^I\over dr}F_I-X^I{d\bar F_I\over dr}\Big)
\label{nat}
.\eq
If we write
\be
i(X^I-\bar X^I)=f^I(r),\qquad i(F_I-\bar F_I)=g_I(r)
\label{decompose}
\eq
Then (\ref{nat}) can be rewritten in the form 
\be
{d\over dr}e^{-2U}=2\Big(X^I{dg_I\over dr}-F_I{df^I\over dr}\Big)
\label{good}
\eq
where we have made use of the relation imposed by the underlying 
special 
geometry,
\be
X^I\partial_rF^I-F^I\partial_rX^I=0,
\label{useful}
\eq
which directly follows from the second relation in  (\ref{berlin}). 

Now comparing (\ref{good}) with
(\ref{gg}) we immediately arrive at the following result
\be
f^I(r)={\tilde h}^I+{p^I\over r}, \qquad g_I(r)={h}_I+{q_I\over r}.
\label{sol}
\eq
where $h_I$ and ${\tilde h}^I$ are constants which must obey 
the constraints coming from demanding the asymptotic flatness of our 
solution as well the condition 
(\ref{ha}). We now turn to show that the solution specified by 
(\ref{s}) and (\ref{sol})
also solves the differential equation (\ref{eqtwo}) obtained from the
vanishing of the gauginos supersymmetry transformation. 

If we multiply both sides of (\ref{eqtwo}) by $f_i^{J}$
and use (\ref{dirac}), then the following equation is obtained.
\be
f_i^{I}{dz^i\over dr}=i{t^I} {e^{U}\over r^2}+
\bar L^I Z{e^U\over r^2},
\eq
from which one can derive the following two relations 
\begin{eqnarray}
{e^U\over r^2}(Z\bar L^I-\bar ZL^I)&=&-ip^I{e^U\over 
r^2}
+2i \hbox{Im}(f_i^I {dz^i\over dr})\label{kinsky}\\
{e^U\over r^2}(Z\bar M_I-\bar ZM_I)&=&-iq_I{e^U\over 
r^2}
+2i \hbox{Im}(h_{iI} {dz^i\over dr})
\label{espen}
\end{eqnarray}
where we remind the reader that 
\begin{eqnarray}
f_i^I&=&(\partial_i +{1\over 2}\partial_iK) L^I\nonumber\\
h_{iI}&=&(\partial_i +{1\over 2}\partial_iK) M_I
.\nonumber
\end{eqnarray}
To evaluate the right hand side of (\ref{kinsky}) 
in terms of the symplectic
sections, we first note the following relation 
\be
\partial_i K{dz^i\over dr}=ie^{K}(\bar F_I\partial_r X^I-\bar 
X^I\partial_rF_I)
\label{pain}
\eq
which upon using  
(\ref{decompose}) and (\ref{useful}), can be rewritten as
\be
\partial_i K{dz^i\over dr}=e^{K}(\bar F_I\partial_r f^I-\bar 
X^I\partial_rg_I)
.\label{pa}
\eq
Using (\ref{pa}) we get
\begin{eqnarray}
f_i^I{dz^i\over dr}& = &(\partial_i +{1\over 2}\partial_iK) 
L^I
{dz^i\over dr}\nonumber\\
& = &\partial_i K{dz^i\over dr}L^I+e^{K\over2}{dX^I\over 
dr}\nonumber\\
& = & e^{K}(\bar F_J\partial_r f^J-\bar 
X^J\partial_rg_J)L^I+e^{K\over2}
{dX^I\over dr}
\label{veryuseful}
\end{eqnarray}
Substituting the relation (\ref{veryuseful}) into equation (\ref{kinsky}) 
gives the following differential equation 
\begin{eqnarray}
{e^{U+{K\over2}}\over r^2}\Big[(F_Jp^J-X^Jq_J){\bar L^I}-
(\bar F_Jp^J-\bar X^Jq_J){L^I}\Big]\nonumber\\
=-ip^I{e^U\over r^2}+{e^K}\Big[L^I(
\bar F_J{df^J\over dr}-\bar X^J{dg_J\over dr})-{\bar L^I}
(F_J{df^J\over dr}-X^J{dg_J\over dr})
\Big]-ie^{K\over2}{df^I\over dr}
\end{eqnarray}
Similar manipulation for (\ref{espen}) leads to 
\begin{eqnarray}
{e^{U+{K\over2}}\over r^2}\Big[(F_Jp^J-X^Jq_J){\bar M_I}-
(\bar F_Jp^J-\bar X^Jq_J){M_I}\Big]\nonumber\\
=-iq_I{e^U\over r^2}+{e^K}\Big[M_I(
\bar F_J{df^J\over dr}-\bar X^J{dg_J\over dr})-{\bar M_I}
(F_J{df^J\over dr}-X^J{dg_J\over dr})
\Big]-ie^{K\over2}{dg_I\over dr}
\end{eqnarray}
The above rather ugly equations can be easily seen to be solved by 
(\ref{s}) and (\ref{sol}).

In deriving our solutions the condition 
({\ref{ha}) was imposed on the holomorphic 
sections. This restricts the allowed values of the harmonic
functions defining the black hole solutions. In addition, the harmonic 
functions are normalised in order 
to obtain asymptotically 
flat solutions. The constraint ({\ref{ha}) is the  
vanishing of the K\"ahler connection $Q_\mu$ of the underlying 
K\"ahler-Hodge manifold which is essential if we were to obtain static 
black hole solutions. 
To explain, we rewrite the K\"ahler connection, 
$$Q_\mu = -{i\over2}\Big(\partial_i K\partial_\mu z^i-
\partial_{i^\star} K\partial_\mu {\bar z}^{i^\star}\Big)$$
in the following form 
\be
Q_\mu=-{1\over2}e^{K}\Big(\bar X^I{\partial_\mu F_I}-{\partial_\mu X}^I
\bar F_I-{\partial_\mu\bar X}^I{F}_I+{X}^I{\partial_\mu \bar F}_I).
\eq
Clearly ({\ref{ha}) follows from $Q_\mu=0.$ In order to find 
the explicit additional constraints on the constants of the harmonic 
functions, it is more 
convenient to express $Q_\mu$ in terms of $f^I(r)$ and $g_I(r)$. 
In terms of these quantities, the K\"ahler
connection takes a very simple form,
\be
Q_\mu =-{e^{K}\over2}(f^I\partial_\mu g_I-g_I\partial_\mu f^I)
\eq
For our solutions, where $f^I(r)$ and $g_I(r)$ are given by 
harmonic functions, the 
vanishing of the K\"ahler connection 
does not impose any restrictions on the electric and magnetic charges. 
However, the 
values of $h_I$ and $\tilde h^I,$ related to the values of the scalars at 
infinity are
constrained by the following conditions
\be
{\tilde h}^I q_I-h_I p^I=0.
\eq
This condition implies that the central charge of the 
supersymmetry algebra must be real for our solution. Therefore, 
the vanishing of the K\"ahler 
connections is related to the reality of 
the central charge. This can be easily seen by noting that for our solutions
\begin{eqnarray} 
Z-\bar Z & = &e^{K\over2}\Big((F_I-\bar F_I)p^I-(X^I-\bar X^I)q_I\Big)
\nonumber\\
&= & ie^{K\over2}\Big({\tilde h}^Iq_I-h_Ip^I\Big)=0.
\end{eqnarray}
Clearly for the static solutions, the central charge has to be real in order
for the differential equation
(\ref{g}) to make sense.

To summarise we have found static BPS black hole solutions for $N=2$ 
supergravity coupled to vector multiplets. These solutions are given by
\begin{eqnarray}
ds^2&=&e^K dt^2-e^{-K}d{\vec x}^2,\nonumber\\
    &=&-{i\over (\bar X^IF_I-X^I\bar F_I)}dt^2-
     {i(\bar X^IF_I-X^I\bar F_I)} {d\vec x}^2,
\end{eqnarray}
where
\be 
i(X^I-\bar X^I) = {\tilde h}^I+{p^I\over r},\qquad 
i(F_I-\bar F_I) ={h}_I+{q_I\over r}
\label{gll}
\eq
and 
\be
{\tilde h}^I q_I-h_I p^I = 0,  \qquad e^K_\infty=1.
\label{general}
\eq
Eq. ({\ref{gll}) provides us with the 
\lq\lq{\it generalised stabilisation equations}" which 
express the values of the moduli at any point in space-time.
\section{Reissner-Nordstr{\o}m Solutions From Special Geometry}
As an application, we use our general solutions and consider the
simplest example, i.e., Einstein-Maxwell 
gravity and derive the extreme Reissner-Nordstr{\o}m 
black hole solution of this theory
using the framework of special geometry. In this case, there are no vector
multiplets and the only scalar present is that of the gravitational 
multiplet which contains the graviphoton. 

To start, consider the spherically symmetric Majumdar-Papapetrou metric
\be
ds^2 =V^{-2}(r)dt^2 - V^{2}(r)(d\vec{x})^2. \qquad
\label{tnt}
\eq
It is well known that 
the source-free Einstein-Maxwell 
equations of motion for the metric in (\ref{tnt}) reduce to Laplace's 
equation for $V(r)$
\be
\nabla^2V(r)=0.
\eq
The Majumdar-Papapetrou black hole solutions for 
Einstein-Maxwell theory are known to admit supersymmetry \cite{gh, tod}.
Therefore, we can imbed Einstein-Maxwell theory in  
$N=2$ supergravity and view the metric (\ref{tnt}) as a solution which 
breaks half of the supersymmetry. 
{From} our previous discussion, the scale 
$V^2$ can be identified with $e^{-K}$. If we describe the 
pure $N=2$ supergravity theory by the 
holomorphic prepotential $F(X^0)=-{i\over 2}(X^0)^2$, we obtain
\be 
V^2=e^{-K}=i(\bar X^0F_0-X^0\bar F_0)=2X^0\bar X^0
\eq 
and thus our black hole solution can be expressed by 
\be
ds^2={1\over 2X^0\bar X^0}dt^2-2X^0\bar X^0 (d\vec x)^2,
\label{sidon}
\eq
with 
\begin{eqnarray}
i(X^0-\bar X^0)& =& {\tilde h}^0+{p^0\over r},\nonumber\\ 
(X^0+\bar X^0)&=& {h}_0+{q_0\over r}
\end{eqnarray}
where $q_0,p^0$ are the electric and magnetic quantum charges 
of the $U(1)$ gauge group associated with the graviphoton. 
Now for a static solution with both charges present, $X^0$ is complex 
and is given by\footnote{An important point is to notice that the 
gauge choice $X^0=1$ is not convenient for the study of black hole 
solutions of $N=2$ supergravity.}
\be
X^0={1\over2}(h_0+{q_0\over r})-i{1\over2}(\tilde h^0+{p^0\over r})
.\eq

The conditions of asymptotic flatness and the vanishing of the K\"ahler
connection gives the following conditions
\be
{\tilde h}^0 q_0-h_0 p^0=0,  \qquad ({\tilde h}^0)^2+(h_0)^2=2.
\eq
This fixes completely the values of $h_0$ and $\tilde h^0$ to
\be
{\tilde h}^0={\sqrt2}{p^0\over\sqrt{p_0^2+q_0^2}}, 
\qquad {h}_0={\sqrt2}{q_0\over\sqrt{p_0^2+q_0^2}}
\eq
and the metric ({\ref{sidon}) can be written as
\be
ds^2 =\Big(1+{1\over r}\sqrt{{p_0^2+q_0^2\over2}}\Big)^{-2}dt^2 -
\Big(1+{1\over r}\sqrt{{p_0^2+q_0^2\over2}}\Big)^2(d\vec{x})^2,\qquad
\eq
which is the extreme Reissner-Nordstr{\o}m black hole solution
of Einstein-Maxwell gravity. 

\section{Entropy and Minimal Central Charge}
In this section, the well known behaviour of static extreme 
black holes at the near horizon is rederived and confirmed using our 
explicit general solutions. 

Near the horizon, the constants ${\tilde h^I}$ and $h_I$ drop out 
and the metric scale $e^{-K_h}$ can be approximated as follows
\begin{eqnarray}
e^{-K_{h}}&=& i(\bar X^I_h F_{Ih}-X^I_h\bar F_{Ih})\nonumber\\
&=& i\Big[(X^I_h+i{p^I\over r})F_{Ih}-X^I_h(F_{Ih}+i{q_I\over 
r})\Big]\nonumber\\
&=& {1\over r}(X^I_hq_I-F_{Ih}p^I)\nonumber\\
&=& -{1\over r} Z_h e^{-K_{h}\over2}.\nonumber
\end{eqnarray}
This implies that the near horizon metric takes the 
Bertotti-Robinson form  
\begin{eqnarray}
ds^2&=&{r^2\over M^2_{BR}} dt^2-{M^2_{BR}\over r^2}(d\vec x)^2\nonumber\\
&=&{r^2\over Z^2_{h}} dt^2-{Z^2_{h}\over r^2}(d\vec x)^2
.\end{eqnarray}

Next, near the horizon, the \lq\lq{\it generalised stabilisation equations}"
expressed in terms of the covariantly holomorphic sections, 
reduce to the following equations
\begin{eqnarray}
i{e^{-{K_h/2}}}(L^I-\bar L^I)_h&=&{p^I\over r}\\
i{e^{-K_h/2}}(M_I-\bar M_I)_h&=&{q_I\over r}
\end{eqnarray}
Using the relation 
$\textstyle{e^{-K_{h}/2}=-{Z_{h}\over r}}$, one obtains 
\begin{eqnarray}
iZ_h(\bar L_h-L_h)&=&{p}^I\\
iZ_h(\bar M_{Ih}-M_{Ih})&=&{q^I}
\end{eqnarray}
Which are the stabilisation conditions found in \cite{fk}.
This also means that, for our solutions, 
the central charge is extremum at the horizon
\be
(D_iZ)_{h}=0.
\eq

Clearly, at the horizon, the central charge (entropy) and the values of the
scalar fields are completely independent of the values of the scalar fields at
spatial infinity which agrees with the results of \cite{fks, s, fk}.

\section{Discussions}
In this work we derived  general $N=2$ static black hole solutions for 
ungauged $N=2$ supergravity theories coupled to an arbitrary number of 
vector and hypermultiplets. The solutions found are spherically symmetric 
Majumdar-Papapetrou like metrics \cite{mp} and are entirely 
expressed in terms of 
the K\"ahler potential of the underlying special K\"ahler manifold spanned 
by the scalars of the vector multiplets. For these solutions,   
the imaginary part of the holomorphic sections are given by a 
set of constrained harmonic functions which depend on the electric and
magnetic charges. This is not surprising since 
the $N=2$ supergravity theory 
can be fully constructed out of the
holomorphic sections. Therefore, one should be able 
to express the black hole 
solutions in terms of symplectic invariants of the 
underlying special K\"ahler manifold. For our static solutions, 
the symplectic invariant is simply the K\"ahler potential. 
This implies that any perturbative or non-perturbative 
corrections to
our black hole solutions can be understood in terms of corrections to 
the K\"ahler potential of the scalar fields in the theory. 

It was also found that the ansatz for the static solutions forces one to set
the K\"ahler connection to zero. Thus one expects 
the Kahler connection (which is also symplectic invariant) to play a 
role in the construction of stationary 
solutions. We will report on the stationary solutions in a separate 
publication.
The role of the K\"ahler connection can be easily seen from 
the following simple observation. 
It is known from the work of Tod \cite{tod} that the most 
general form of the metric admitting 
supersymmetries can be written in the form 
\be
ds^2={V\bar V}(dt+\vec w.\vec dx)^2-({V\bar V})^{-1}(d{\vec x})^2
\eq
where (in the absence of dust), $V$ is the inverse of a harmonic 
function and $w$ is 
defined by 
\be
\vec\nabla\times \vec w=-{i\over(V\bar V)^2}(\bar V\vec\nabla V-
V\vec \nabla \bar V).
\eq
These solutions constitute a class of stationary metrics known as 
\lq\lq{\it conformastationary}" 
and were discovered by Neugebauer \cite{neu}, Perj\'es \cite{pe}
and Israel and Wilson \cite{iw}. To see how the symplectic 
invariant K\"ahler connection is 
related 
to $w$, we note that if we write $e^K=V\bar V,$ with $V$ time 
independent, then 
the Kahler connection vector becomes
\begin{eqnarray}
\vec Q&=&-{i\over 2V{\bar V}}(\bar V\vec\nabla V-V\vec \nabla \bar V)\nonumber
\\&=&{1\over2}(V\bar V)(\vec\nabla\times \vec w).
\end{eqnarray}

Finally, we mention that while the constraints on 
the constants 
$(h_I, \tilde h^I)$ fix them 
completely for the case of pure Einstein-Maxwell gravity, 
in the presence of vector multiplets one 
still have freedom in choosing these constants. 
This allows one to study massless black holes 
configurations and the interplay between space-time and moduli-space 
singularities \cite{coni, gms}. 
This subject is currently under investigation. 
\section*{Acknowledgements}
This work is supported by DFG and in part by DESY-Zeuthen. 
I thank D. L\"ust, S. K. Espenlaub, T. Mohaupt, S. Mahapatra and 
A. Van Proeyen for conversations and special thanks to K. Behrndt for 
many useful discussions.  
\newpage


\begin{thebibliography}{100}

\bibitem{fks} S. Ferrara, R. Kallosh, A. Strominger, {\it Phys. Rev.}
{\bf D52} (1995) 5412.

\bibitem{s} A. Strominger, {\it Phys. Lett.} {\bf B383}
(1996) 39.

\bibitem{fk} S. Ferrara, R. Kallosh,  {\it Phys. Rev.}
{\bf D54} (1996) 1514.

\bibitem{ksw} R. Kallosh, M. Shmakova and W.K. Wong,  
{\it Phys. Rev.}
{\bf D54} (1996) 6284. 

\bibitem{g} K. Behrndt, R. Kallosh, J. Rahmfeld, M. Shmakova, W.K. 
Wong, {\it Phys. Rev.} {\bf D54} (1996) 6293. 

\bibitem{r}G. Lopes Cardoso, D. L\"ust and T. Mohaupt, 
{\it Phys. Lett.} {\bf B388} (1996) 226. 

\bibitem{qe}K. Behrndt, G. Lopes Cardoso, B. de Wit, R. Kallosh, D. L\"ust 
and T. Mohaupt, {\tt hep-th/9610105}.

\bibitem{o}S. -J. Rey, {hep-th/9610157}.

\bibitem{u} W.A. Sabra, {\it Mod. Phys. Lett.} {\bf A12} (1997) 789, 
{hep-th/9611210}. 

\bibitem{p}M. Shmakova,{hep-th/9612076}.

\bibitem{ber} K. Behrndt, hep-th/9610232.
\bibitem{me} K. Behrndt and W. A. Sabra, hep-th/9702010, to appear in 
{\it Phys. Lett.} {\bf B} (1997).
\bibitem{mea} W. A. Sabra, hep-th/9703101.
\bibitem{fkg} S. Ferrara, R. Kallosh and G. W. Gibbons, hep-th/9702103
\bibitem{fre}P. Fr\'e, {\it Nucl. Phys.} {B} (Proc.
Suppl.) {\bf 45B},{\bf C} (1996) 59;\\
 P. Fr\'e and Soriani, 
The N=2 Wonderland: From
Calabi-Yau Manifolds To Topological Field Theories, World Scientific (1995);\\
L. Andrianopoli, M. Bertolini, A. Ceresole, R. D'Auria, 
S. Ferrara, P. P. Fr\'e and T. Magri, hep-th/9605032.

\bibitem{G} G.W. Gibbons, in: {\it Supersymmetry, Supergravity and 
Related Topics}, World Scientific, Singapore 1985); 

\bibitem{bh} J. Bekenstein, {\it Lett. Nouv. Cimento} {\bf 4} (1972) 737, 
{\it Phys. Rev.} {\bf 
D7} (1973) 2333; {\it Phys. Rev.} {\bf D9} (1974) 3292; 
S. Hawking, {\it Nature} {\bf 248} (1974) 30, 
{\it Comm. Math. Phys.} {\bf 43} 1975.
\bibitem{vp} B. Craps, F. Roose, W. Troost and A. Van Proeyen, 
{\it What is special K\"ahler geometry?},  hep-th/9703082. 
\bibitem{CDFP}
A. Ceresole, R. D'Auria, S. Ferrara and A. van Proyen,
{\it Nucl. Phys.} {\bf B444} (1995) 92.
\bibitem{special}S. Ferrara and A. Strominger, in Strings
'89, ed. R. Arnowitt, R. Bryan, M. J. Duff, D. Nanopulos and C. N. Pope,
World Scientific, Singapore, (1990) 245.
\bibitem{c} P. Candelas and X. de la Ossa,
{\it Nucl. Phys.} {\bf B355} (1991) 455;
P. Candelas,  X. de la Ossa, P. Green and L. Parkers,
{\it Nucl. Phys.} {\bf B359} (1991) 21.
\bibitem{cc} A. Strominger, {\it Commun. Math. Phys.} {\bf 133}
(1990) 163.
\bibitem{ccc} R. D'Auria, 
S. Ferrara and P. Fr\`e, {\it Nucl. Phys.}
{\bf B359} (1991) 705.
\bibitem{cccc}L. Castellani, R. D'Auria, 
and S. Ferrara, {\it Class. Quantum Grav.}
{\bf 1} (1990) 317.
\bibitem{sspecial}B. de Wit and A. Van Proeyen, {\it Nucl. Phys.}
{\bf B245} (1984) 89; 
E. Cremmer, C. Kounnas, A. Van Proeyen, J. P. Derendinger, S. Ferrara, 
B. de Wit and L. 
Girardello, {\it Nucl. Phys.} {\bf B250} (1985) 385; 
B. de Wit, P. G. Lauwers and A. Van Proeyen, 
{\it Nucl. Phys.} {\bf B255} (1985) 569; S. Cecotti, 
S. Ferrara, and L. Girardello, {\it Int. J. Mod. Phys.}
{\bf A4} (1989) 2475. 
\bibitem{mc}
M. Cveti\v c and D. Youm, {\it Phys. Rev.} {\bf D53} (1996) 584;\\
M. Cveti\v c and A. A. Tseytlin, {\it Phys. Rev.} {\bf D53} (1996) 5619;\\
R. Kallosh and B. Kol, {\it Phys. Rev.} 
{\bf D53} (1996) 5344;\\
M. Cveti\v c and C. M. Hull, {\it Nucl. Phys.} {\bf B480} (1996) 296.
\bibitem{gh} G.W. Gibbons and C.M. Hull, {\it Phys. Lett.} {\bf B109}
(1982) 190. 
\bibitem{tod} K. P. Tod, {\it Phys. Lett.} {\bf B121} 1983) 241;
{\it Class. Quantum Grav.}{\bf 12} (1995) 1801

\bibitem{neu} G. Neugebauer, Habilitationschrift, Jena (1969).

\bibitem {pe}Z. Perj\'es, {\it Phys. Rev. Lett.} {\bf 27} (1971) 1668.

\bibitem{iw}W. Israel and G. A. Wilson, {\it J. Math. Phys.} 
{\bf 13} (1972) 865.

\bibitem{mp} S. D. Majumdar, {\it Phys. Rev.} {\bf 72} (1947) 930;
A. Papapetrou, {\it Proc. Roy. Irish. Acad.} {\bf A51} (1947) 191.

\bibitem {br}B. Bertotti, {\it Phys. Rev.} {\bf 116} (1959) 1331; I. Robinson,
{\it Bull. Acad. Polon. Sci.} {\bf 7} (1959) 351.

\bibitem{fr} P. Fr\'e, hep-th/9701054;
\bibitem{gz} M.K. Gaillard and B. Zumino, {\it Nucl. Phys.} {\bf B193}
(1981) 221.
\bibitem{sym}P. Fr\'e and P. Soriani, {\it Nucl. Phys.} {\bf B371} 
(1992){659}; 
W. A. Sabra, {\it Nucl. Phys.} {\bf B486} (1997) 629.
\bibitem{coni}A. Strominger, {\it Nucl. Phys.} {\bf B451} (1995) 96.
\bibitem {gms}B. R. Greene, D. R. Morrison and A. Strominger, 
{\it Nucl. Phys.} {\bf B451} (1995) 109.
\end{thebibliography}
\end{document}